\documentclass[12pt]{iopart} 
\pdfoutput=1
\usepackage{graphicx} 
\usepackage[caption=false]{subfig}
\usepackage{color}
\usepackage{float,amssymb}

\begin{document} 

\title[]{ 
Entropy and hierarchical clustering:
characterising the morphology  of the urban fabric 
in different spatial cultures
}

\author{  E. Brigatti$^{1*}$, V. M. Netto$^{2*}$, F.N.M. de Sousa Filho$^{3}$ , and C. Cacholas$^2$}

\address{$^{1}$ Instituto de F\'{\i}sica, Universidade Federal do Rio de Janeiro, 
Av. Athos da Silveira Ramos, 149,
Cidade Universit\'aria, 21941-972, Rio de Janeiro, RJ, Brasil.}
\address{$^2$ Graduate Programme in Architecture and Urbanism, Universidade Federal Fluminense,
R. Passo da P\'atria 156, 
24210-240, Niteroi, RJ, Brasil.}
\address{$^3$ Instituto de Computa\c c\~ao, Universidade Federal do Rio de Janeiro,
Av. Athos da Silveira Ramos, 274
Cidade Universit\'aria,
21941-916, Rio de Janeiro, RJ, Brasil.}

\address{* The authors contributed equally to this work.}

\begin{abstract} 

In this work, we develop 
a general method for estimating the Shannon entropy of a bidimensional sequence based on the extrapolation of block entropies. 
We apply this method to analyse the spatial configurations of cities of different cultures and regions of the world.
Findings suggest that this approach can identify similarities between cities, generating accurate results for recognising and classifying different urban morphologies.
The hierarchical clustering analysis based on this metric also opens up new questions about the possibility that urban form can embody characteristics related to different cultural identities, historical processes and geographical regions.
\end{abstract}

\maketitle

\section{Introduction}

The characterisation of the morphology of cities
is a vast area of investigation
which traditionally interests a wide range of disciplines, including 
urbanism, architecture and geography.
Around the 1970s, some innovative 
attempts, based on the introduction of 
quantitative approaches, have deeply impacted a long-established practice
which, before of that, was focused on more qualitative speculations and descriptive 
procedures.
Among the new concepts introduced by these studies, entropy has proven 
to be a very fertile, though elusive, idea which has generated different and successful
methodologies. 

The establishment of an analogy of the thermodynamic concept of entropy 
applied to geographical systems suggested a new approach for characterising 
and describing the evolution of the state of an urban system \cite{analogy}.
Building up on this approach, maximum-entropy methods \cite{maxent} 
have been  introduced to model the urban dynamics and its forms \cite{maxentUrban,maxentUrban2}.
Finally, entropy measures have been used for characterising patterns 
of spatial systems \cite{walsh,patterns1,patterns3,nos3,nos4}.
In this work we will focus on this last aspect.

In a wide variety of studies, entropy has been used just as a diversity index:
a quantitative measure that reflects how equiprobable
are some specific different types 
of a local spatial character used to describe a dataset.
In fact, entropy evaluation of a given probability distribution is
a unique and unambiguous criterion \cite{Jaynes57} 
for quantifying the  intuitive notion of difference between a broad and a sharply peaked distribution.
This notion can be naturally related to a 
general idea of uncertainty and randomness,
where larger entropies correspond to broader distributions closer to equiprobability,
and so, randomness.
This formulation was originally proposed by Claude Shannon,
in a richer context, where the considered dataset were strings 
of text or, more generally, sequences of information carrying symbols \cite{Shannon48}. 
In this setting, 
Shannon's  entropy quantifies the uncertainty 
associated with predicting  a letter which follows a well known portion of a string.
In this scenario, Shannon's entropy is no more just a general diversity index but becomes 
rigorously connected to the notion of information, in a well defined 
signification: the degree of surprise the 
source that produces the sequence causes on 
the observer.
 
The simple use of entropy as a diversity index can be found in a broad amount of works, 
based on the election of a
specific local trait characterising the spatial morphology. 
Some recent  
examples can be found in the orientations \cite{Gudmundsson13,Boeing19} 
and lengths \cite{Gudmundsson13} of streets and
the parcel sizes probability distributions \cite{Bitner21}. 

A quite different situation corresponds to consider 
the built environment as a whole, not selecting 
just a single local trait for characterising it.
In 
such an approach, an informative representation of the complete urban form
must be determined. 
Maps illustrating the two-dimensional arrangements 
based on building footprints are a possible choice.
Figure-ground diagrams, or Nolli maps, 
are a classic methodological resource in urban studies.
By defining these cellular arrangements, we capture the structures of urban blocks in relation 
to the open spaces of streets and public squares.
Street networks and blocks
are clearly a by-product or a more coarse-grained
representation extracted from these maps. In this sense, they carry a more 
partial view of the spatial system. 

In our analysis, built form maps 
were 
converted in a matrix of binary values 
which we consider as an information carrying image,
where the symbols correspond to built or unbuilt spaces. 
Our goal is to measure the Shannon entropy of these 
two-dimensional representations. 
Note that previous 
attempts to explore these ideas 
were not able to connect to 
Shannon's work
because they simply 
estimated low order entropies: they measured the probability distribution
of a single symbol or, at most, of symbol-pairs 
of the entire image \cite{Nowosad19}.
In contrast, to measure the Shannon's entropy of the global system, and not just grasp 
local properties of the map, the distribution probability of the state of the 
entire image must be quantified. 
Even if it seems a paradox, as, in general, we have one image and not an ensemble 
of them, for one dimensional strings a vast toolkit of solutions has been used since the 
1990s \cite{schurmann1996entropy}. 
The extension of these approaches to two dimensional systems, 
e.g. images, followed a more 
hesitant path \cite{Grassberger86}, but different examples appeared 
more recently \cite{Feldman2003, Feldman_Chaos08}.

Here we elaborate over this theory, estimating, as a first step, the Shannon entropy of Nolli maps 
of different cities of the world.  
As maps 
are represented by matrices 
of 1 and 0,  
theoretically, this  
procedure corresponds to measuring
the Shannon 
entropy of a two-dimensional binary symbolic sequence.
As mentioned, in this context, Shannon entropy has a precise interpretation in terms
of 
information: 
it is a measure of the surprise a source that produces the sequence causes in the observer \cite{Shannon48}. %
Physical arrangements marked by higher levels of randomness, and so by 
higher entropy levels
are characterised by a greater unpredictability. 
In contrast, the presence of  patterns and regularities in urban structures 
corresponds to lower entropy, which means a higher predictability.

In the spatial literature it is 
common to find  a concern about directly 
measuring entropy in spatial data \cite{patterns3}. 
This concern is genuine only when it is considered 
not the entropy of the system but, naively, the entropy of a single symbol, which is obviously aspatial.
In turn, our
approach is certainly  
measuring the entropy of the entire system, taking into account 
all spatial correlations 
and estimating it at all scales. 
Moreover, this measure reflects an authentic global character of the map,
as global features 
of configurations are estimated. 
This is quite different from other classic 
approaches, which calculate mean values of local
measures and consider them as global properties. 
Entropy measure is more sensitive and general than traditional 
two-point measures, like standard correlations. These measures, 
are not able to distinguish correlations that differ over more than two points
and are sensitive over a fixed, single, spatial period. These aspects contrast with 
our approach, which is based on an unparameterized 
function of the distribution of the
spatially-extended configurations. 

In the following, we develop an in-depth 
analysis of this methodology,
testing different approaches and the robustness of the most 
sensitive aspects of the method.
Our aim is to make use of this entropy estimation 
for analyzing the spatial configurations of the urban fabric of different cultures and regions of the world,
with the hope that this methodology can identify similarities between 
cities, generating accurate results for recognition and classification of the urban morphologies.
Our dataset contains cities that 
present a significant variance in the
density of built form cells and these differences can influence 
significantly the entropy value, not being necessarily connected with the
randomness  and correlations 
of the built form. For this reason, we introduce 
a new heuristic approach for correcting this effect
and allowing a more proper comparison of
urban systems with significant difference in their 
occupation densities.
The estimation of entropy allows us to develop 
a precise measure of the randomness of the system.
As, more in general, we are interested in characterizing the structures and 
patterns of our system, which are the product of the effect of correlations, 
we also estimate the excess entropy,  a well-defined measure
of statistical complexity \cite{Grassberger86,excess,excess2,excess3}.

The systematic application of this methodology
allows the analysis of a vast dataset, with urban sections 
from 68 cities in six continents.
Once we manage to obtain 
the entropy-based measures of these 
sections, 
we shall use 
these results for defining a similarity measure to compare and cluster the studied
cities. 
Finally, 
we will interpret how the obtained hierarchical clusterisation
can be associated with specific regions or spatial cultures, highlighting intuitive interpretations of the results
or reasons of non-contingent similarities or unexpected differences between cities.

\section{Materials}

As a first step, the urban form is reduced to two-dimensional arrangements based on building footprints.
Since Giambattista Nolli's 1748 Map of Rome, the figure/ground diagrams have become a 
methodological resource in urban studies, showing built/unbuilt distinctions \cite{verstegen2013giambattista}. 
We collect our set of empirical cases
selecting cities on the basis of
their importance in their region or country and
the availability of information on built form. 

For methodological reasons, we select areas within these cities 
following two critical considerations. 
The first 
observes that it is interesting to decouple the analysis of urban structures between small-scale, detailed and denser urban areas, and large-scale regional and peripheral urban areas.  
In fact, such 
areas are different from each other, and for this reason, 
they can be naturally described using different methodologies. 
The first small-scale urban area is defined by specific features such as buildings and urban blocks,
which introduce 
characteristic scales.
This means that there are some well-defined scales related to the distance above which 
configurations loose their correlations. These characteristic scales 
define sub-systems distinguished by typical local patterns (urban blocks, 
individual buildings and possible neighbourhoods).
Here, human action is the principal vector 
defining shapes and patterns,  
which generally appear in a stratified form, like the ones we see in older and traditional central areas. 
In turn, large-scale regional and peripheral urban areas are likely to include sparse occupations, frequently with a scale-free character. 
This means that the characteristics of their patterns are independent of the scale we fix for analysing them:
looking at different scales, the underlying structure remains the same.
In these regions, physical features linked to topography, geographical formations and barriers (e.g. water bodies, mountains, and valleys), along with the presence of very large infrastructures (e.g. highways) might play 
relevant roles in the definition of the spatial patterns.
In this work, we will focus only on small-scale areas with dense 
urban form.

The second consideration takes into account that our method is well fitted for estimating 
entropy for 
continuous 
urban areas. 
The high continuity and homogeneity
of built form allows us to use a specific extrapolation technique that will prove useful for estimating the entropy of our two dimensional symbolic sequences.
For these reasons, the selection of  sections was based on the identification of 
areas with a high spatial continuity in the fabric of built form. 

\begin{figure}[h]
\centering
\includegraphics[width=0.8\textwidth, angle=0]{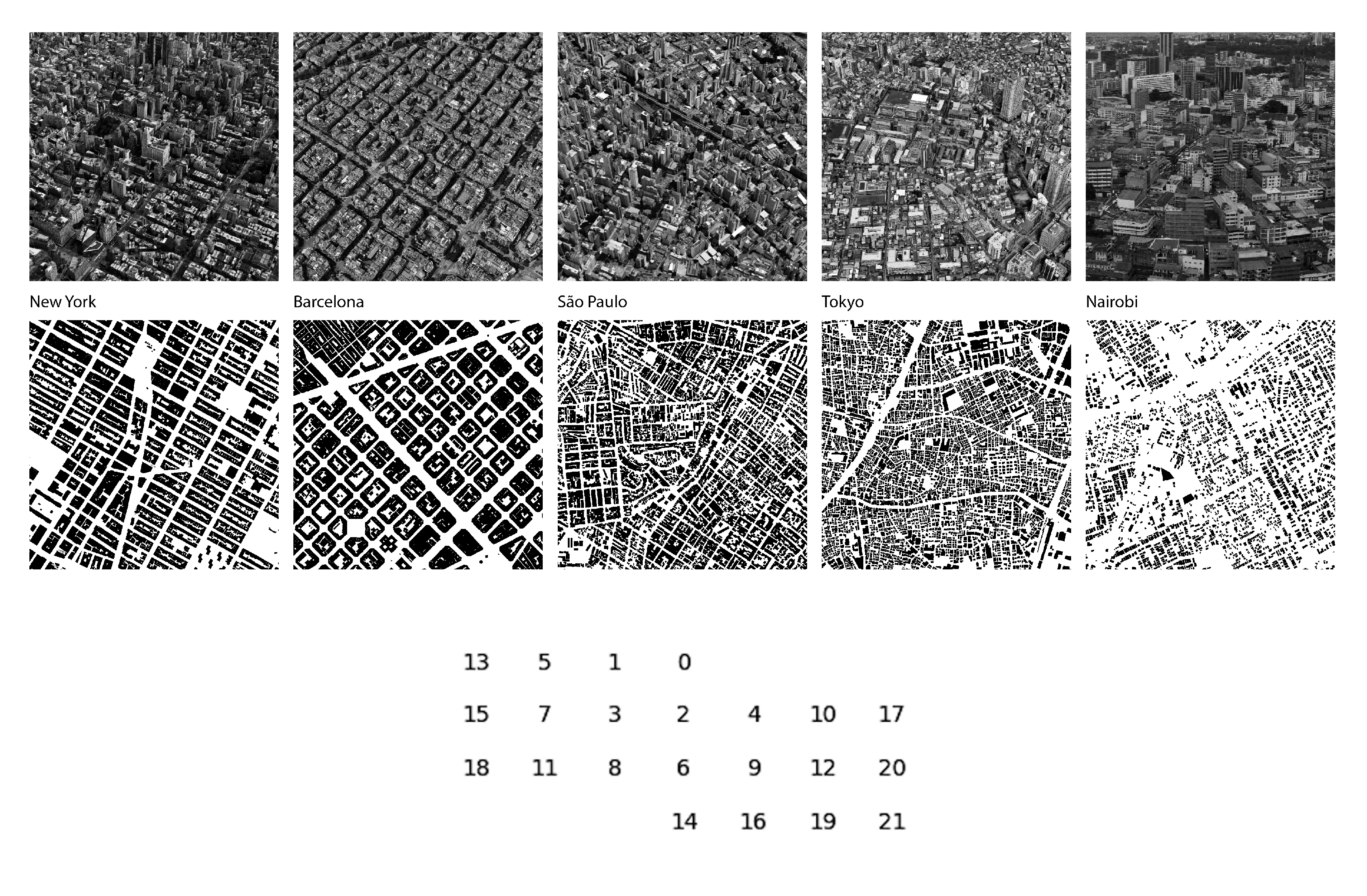}
\caption{ {\it Top:}  Urban sections with the corresponding Nolli's maps with building footprint 
in downtown areas of some exemplar cities. 
{\it Bottom:} the 2D blocks used for evaluating the differential
entropies $h_n=H_n-H_{n-1}$.}
\label{fig_data}
\end{figure}

We prepared our sample extracting building footprints in sections of cities from the public map 
repository Google Maps API. 
We tested trade-offs between resolution and data availability 
for distinct scales. 
We chose geographic areas of $9,000,000$ m$^2$, which were considered sufficient for 
representing the general spatial characteristics of 
small-scale urban areas 
of 68 cities around the world (figure \ref{fig_data}).
Built form maps were exported in high resolution, filtering layers and converting entities representing buildings into solid raster cells. 
Images underwent a re-sizing process 
and were converted to a monochrome system and then into a matrix of size $1000\times 1000$ cells with binary numerical values (figure \ref{fig_data}).\\

\section{Methods} 

Estimation of the Shannon entropy of the considered two dimensional (2D) cellular arrangements uses a method commonly applied for estimating the entropy of sequences of symbols encoded in one-dimensional (1D) strings \cite{schurmann1996entropy}.
For 1D data sets, the method consists of defining the block entropy of order $n$ through

\begin{equation}
H_n=-\sum_k  p_n(k) \log_{2}[p_n(k)],
\label{1entropy}
\end{equation}
where blocks are string segments of size $n$, and the  sum runs over all the $k$ possible $n$-blocks.
Equation~(\ref{1entropy}) corresponds to the Shannon entropy of the probability distribution $p_n(k)$.
The Shannon entropy 
of the considered system (the whole 1D string) \cite{schurmann1996entropy,lesne2009entropy}, which we indicate with $h$, is obtained from the limit:

\begin{equation}
h=\lim_{n \to \infty} H_n/n,
\label{hentropy}
\end{equation}

which measures the average amount of randomness per symbol that persists after all correlations and constraints are taken into account.
The above limit exists for all spatial-translation invariant systems, as demonstrated in \cite{cover1991elements}.
Equation \ref{hentropy} gives precisely the entropy for a theoretical infinite set of data.
In real situations with finite data set, 
the method estimates the probabilities of distinct arrangements of cells within blocks up to a certain size $n$, counting their frequencies. 
Alternatively, the Shannon entropy can be evaluated as the limit of the differential
entropies $h_n=H_n-H_{n-1}$: 
\begin{equation}
h=\lim_{n \to \infty} h_n,
\label{hentropy2}
\end{equation}
(note that, for definition $H_0=0$).
This is the limit of a form of conditional entropy, as $h_n$ 
is the entropy of a single symbol 
conditioned on a block of $n-1$ adjacent symbols \cite{Feldman2003}. 
The two limits (Eqs. \ref{hentropy} and \ref{hentropy2}) are equivalent. 
More details about these methods can be found in \cite{schurmann1996entropy,Feldman2003,lesne2009entropy}.

These approaches can be generalised to sequences of symbols in two dimensions  
by defining the $n$-blocks for a 2D matrix \cite{Feldman2003}. 
For the method of equation \ref{hentropy}, 
we implemented  
in a previous work \cite{nos4} the intuitive idea of considering a block of size $n$ as a square which contains $n$ cells. 
We obtained the sequence of $H_{n}$, also for $n$ values that do not correspond to squares, considering blocks that interpolate these perfect squares.
Here we generalise this approach scanning the matrix with random paths. 
Once fixed a starting point, the following points  
are selected from the cells on the outer edge,
among its first, not previously visited, neighbours.
This simple rule generates any considered block of size $n$. 
In contrast, for the case of equation \ref{hentropy2}, we used the 2D blocks defined by Feldman {\it et al.} in \cite{Feldman2003} and depicted in Fig. \ref{fig_data}.

The advantage of using the approach in 
equation \ref{hentropy}
is that the set of  $H_n/n$ values is monotonous 
and concave and, in general, displays a clear regularity.
For this reason, the limit 
can be empirically obtained fitting the 
$H_n/n$ points with an appropriate function
and then taking its limit for $n \to \infty$.
We found heuristically that, for all examined cases and independently of the random paths
used for scanning the matrix, the following ansatz provides an excellent fit:
\begin{equation}
H_n/n \approx a+c/n^b, \qquad  b,c>0.
\label{fitting}
\end{equation}
For a given random path $i$ which defines the blocks, the fitted value of 
$a$ gives a reasonable extrapolation of the Shannon Entropy $h_i$. 
To 
assure the independence of the method on the particular path, 
we repeated the algorithm selecting 100 different randomly generated paths. 
The value of $h$ can be estimated by using 
the mean value 
of the different $h_i$. 

In contrast, generally, the second approach in 
equation \ref{hentropy2}
presents a faster convergence, but it is more 
influenced by statistical errors \cite{schurmann1996entropy}.
\\


From these measurements we obtain the value of the Shannon entropy 
of the considered maps. 
Our goal is to develop a classification scheme based on the differences 
between these values. 
Our dataset contains cities with varying
density of built form cells.  If the density value is far from 50\%,
there is an important reduction 
in the entropy value caused only by
this asymmetry. This fact is not necessarily connected with the
randomness  and correlations 
of the built form.
For example, the presence of a river, which corresponds to 
a uniform unbuilt region, can reduce sensibly the entropy value. 
As we are interested in characterising the randomness 
of the built form,  
we introduce an heuristic approach for tackling this problem: 
we correct the entropy value $h$ 
adding the term which corresponds to the reduction 
of the entropy value due to the
frequency of 0 and 1 present in the data set:

$h_c= h+(1-H_1)$.

This procedure does not correct all the contributions
that an asymmetry in the frequency of 0 and 1 have on the value 
of $h$ but, at least, it takes away the larger ones. 
In fact, the principal influence of the symbol distribution on the entropy value
is encapsulated in $H_1$. 
The way $h$ differs from $H_1$ captures in an integrated and involved form
how the presence of correlations determines the effective randomness
\cite{lesne2009entropy}.\\


The estimation of entropy 
gives a precise measure of the randomness or unpredictability of the system.
If we are interested in  characterising  the structures and 
patterns of our system,
alternative quantities must be considered.
A well defined and interesting one is the excess entropy  \cite{Grassberger86,excess}, 
which is obtained by examining how the finite entropy  estimates $h_n$ (equation \ref{hentropy2}) 
converge to their asymptotic value $h$. 
There are different definitions for the excess entropy in 2D \cite{Crutchfield_Chaos03},
and a possible one is obtained in terms of block entropies.
If the system is scanned considering only blocks of size up to $n$, 
the system appears to have an entropy of $h_n$. This means that 
 the system appears more random than it actually is by an amount $h_n-h$. 
 By summing up these entropy overestimates, 
we may obtain the excess entropy
 \cite{Grassberger86,excess}: 
 
\begin{equation}
E=\sum_{n=1}^{\infty} h_n-h.
\label{excess1}
\end{equation}
The excess entropy thus measures the amount of apparent
randomness 
that is  recognised as a regularity if looked at a larger scale,
where new correlations appear. 
For this reason,  
the excess entropy can be considered a measure of the global structure of the system.
For many cases of interest this sum is not finite
and it is interesting to estimate the finite-n expressions for 
$E$ \cite{Crutchfield_Chaos03}: 
\begin{equation}
E_M=\sum_{n=1}^{M} h_n-h_M.
\label{excess2}
\end{equation}
where $M$ is a finite and fixed value.
\\

Our empirical results will show that $h_c$ is the best marker
for realising the classification scheme of our dataset, performing 
clearly better than $h$ or $E$.
We quantify the levels of similarity between cities by using the estimated corrected entropies $h_c$.
The corresponding dissimilarity measure is obtained
defining a distance  between cities $i$ and $j$ based on the values of $h_c$:
$d^{ij}=|h_c^i-h_c^j|$. 
We created a matrix of distances for all the analysed cities and then defined a network where cities are nodes, and edges (links between nodes $i$ and $j$) are present only if the value of $d^{ij}$ is smaller than a fixed threshold value. 
The detection of communities displayed by these networks is realised by using 
the Louvain method described in \cite{louvain}, 
implemented in the Python module {\it community}  \cite{community} 
which depends on the NetworkX python package.

We further developed the cluster analysis constructing a dendrogram representation of the distance matrix
\cite{barbrook1998phylogeny,benedetto2002language}. 
We used the unweighted pair group method with arithmetic mean (UPGMA). This method shapes a dendrogram that reflects the structure present in the similarity matrix, building a hierarchy of clusters. 
The algorithm used in the analysis is part of the module Bio.Phylo in the Biopython package \cite{biopython}.

\section{Results}

\begin{figure}[h]
\centering
\includegraphics[width=0.45\textwidth, angle=0]{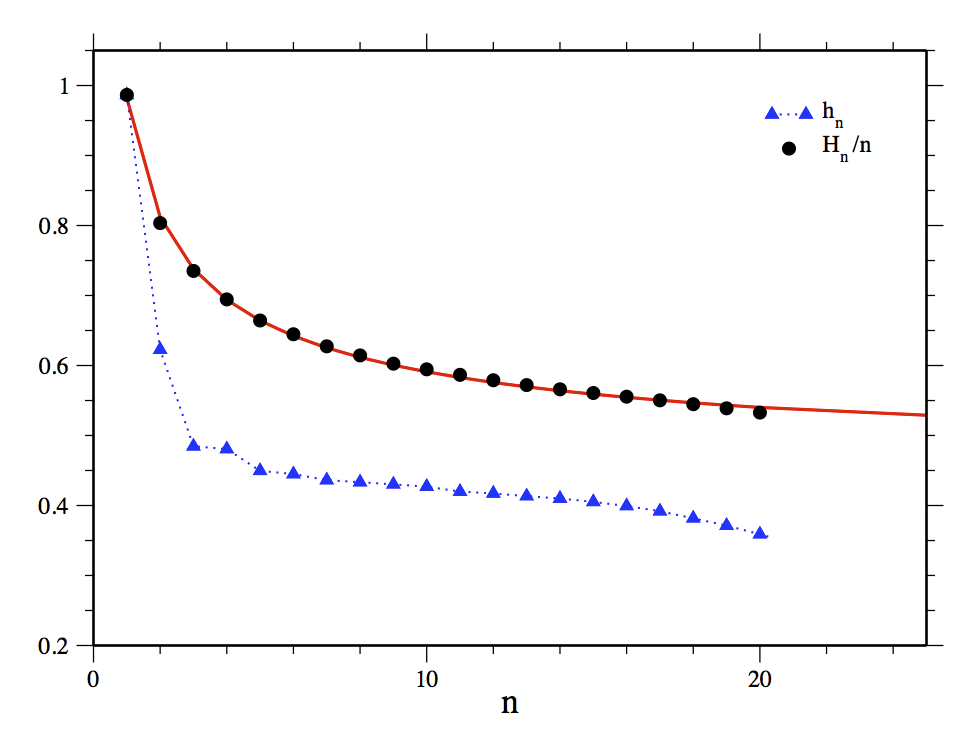}
\includegraphics[width=0.45\textwidth, angle=0]{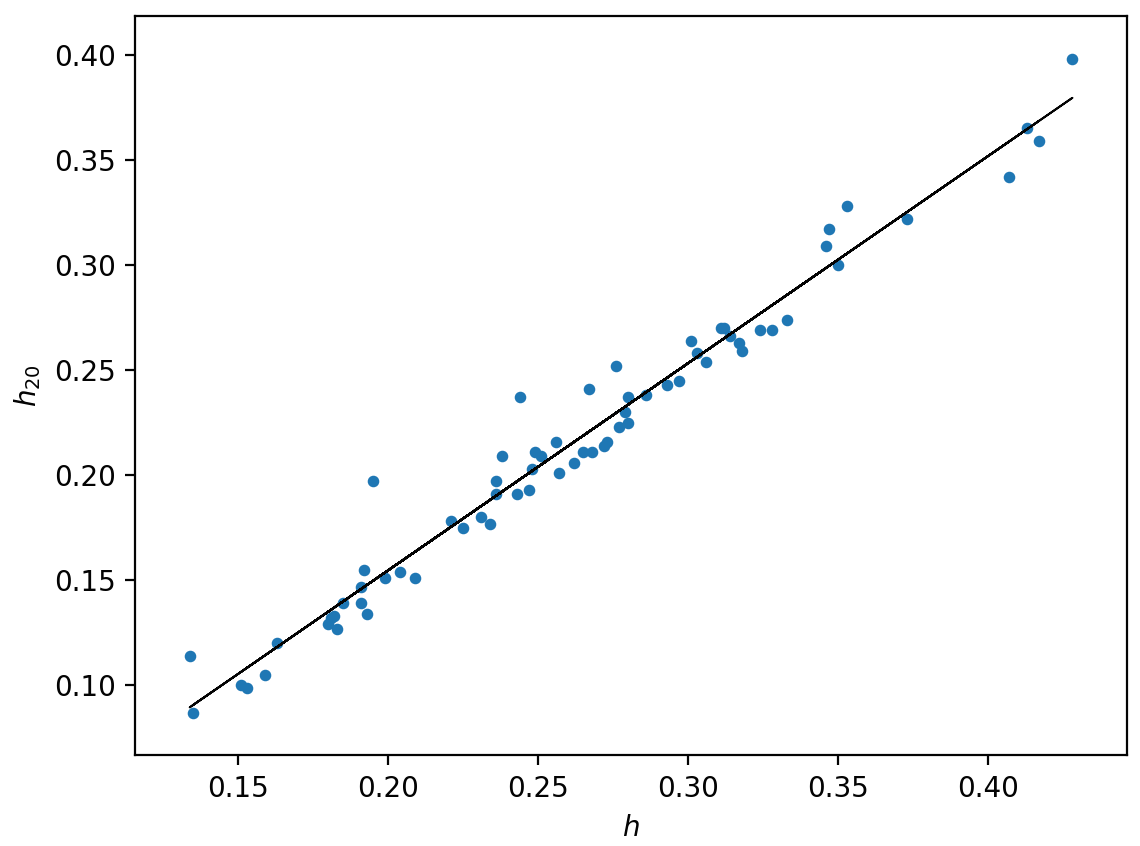}
\includegraphics[width=0.95\textwidth, angle=0]{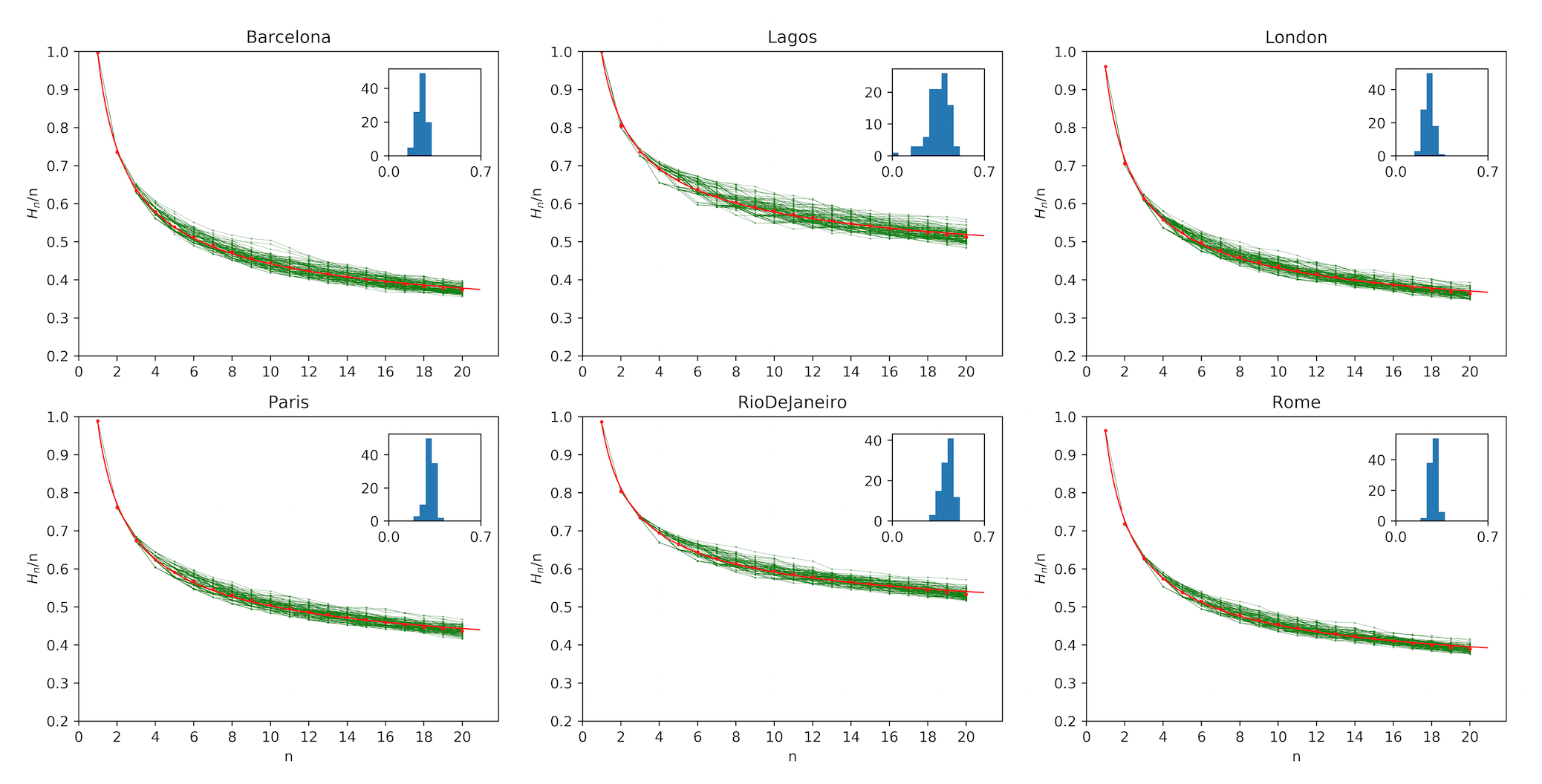}
\caption{ {\it Top:} On the left we plot, for the city of Rio de Janeiro, 
the estimated values of $H_n/n$  for a specific random path. 
The same figure displays the values of $h_n$. 
The continuous red line represents the best fitting of the $H_n/n$ data using the function in eq. \ref{fitting}.
The discontinuous blue line is a guide for the eyes.  
All the analysed cities present analogous behaviour.
On the right we show the scattering plot for the $h$ values of all the considered cities 
obtained by 
extrapolating the $H_n/n$ values or based on the estimation of $h_{20}$.
The continuous line represents the linear regression $y=0.99\cdot x-0.04$. 
{\it Bottom: } the estimated values of $H_n/n$  for 100 different random paths. 
Red points are their mean values 
and the continuous red line is the best fitting obtained using the eq. \label{fitting}.
In the inset, the distribution of the $a_i$ values 
for all the considered random paths.  
Data are shown for six paradigmatic cities.
}
\label{fig_entropy}
\end{figure}

Figure \ref{fig_entropy} shows the results for the estimation of $H_n/n$ and $h_n$
for the city of Los Angeles. This is a paradigmatic example of the general behaviour 
of our approach.
We can note how the set of  $H_n/n$ values, here presented for a specific random scanning path, 
displays a regular, monotonous and concave shape.
For this reason it is possible to use an empirical function for fitting the points and estimating the limiting behaviour for $n\to \infty$. The use of equation \ref{fitting} provides an excellent fit.
Despite the relative slow convergence, the fine quality of the fits allows a good extrapolation of the Shannon Entropy $h$.
The same figure displays the estimation of $H_n/n$ for 100 different 
scanning paths. For each $i$ path we fit equation \ref{fitting} and estimate the $a_i$ value. 
We calculate the mean value $a$ from the distributions of the $a_i$, which show well behaved unimodal shapes with relative small variances. 
Alternatively, a general $a$ value can be evaluated by fitting equation \ref{fitting}
to the mean values of the $H_n/n$. 
No significant differences are found, but the first procedure is more robust and more grounded theoretically.

The differential entropies approach (Eq. \ref{hentropy2})
shows a faster convergence, but it is more influenced by statistical errors and presents irregularities. 
In particular, the curve is not concave.
We can not use any numerical approximation for evaluating its limiting value and we must estimate 
the value of $h$ using the larger $h_n$, which is $h_{20}$.

The scattering plot for the $h$ values of all the considered cities obtained from the extrapolation of the $H_n/n$ values or based on the estimation of $h_{20}$ (see Figure \ref{fig_entropy})
shows how the two results are perfectly congruent. Considering that the approach that uses
$H_n/n$ 
effectively estimates the $n\to \infty$ limit, and 
it is not dependent on the scanning paths, we use it as the best estimate of $h$.
It is interesting to note that the independence of our estimations of $h$ on the scanning paths 
proves their independence from the 
rotational transformation of our data, a fundamental property of a correct numerical estimation of the Shannon entropy in 2D.

\begin{figure}[h]
\centering
\includegraphics[width=0.9\textwidth, angle=0]{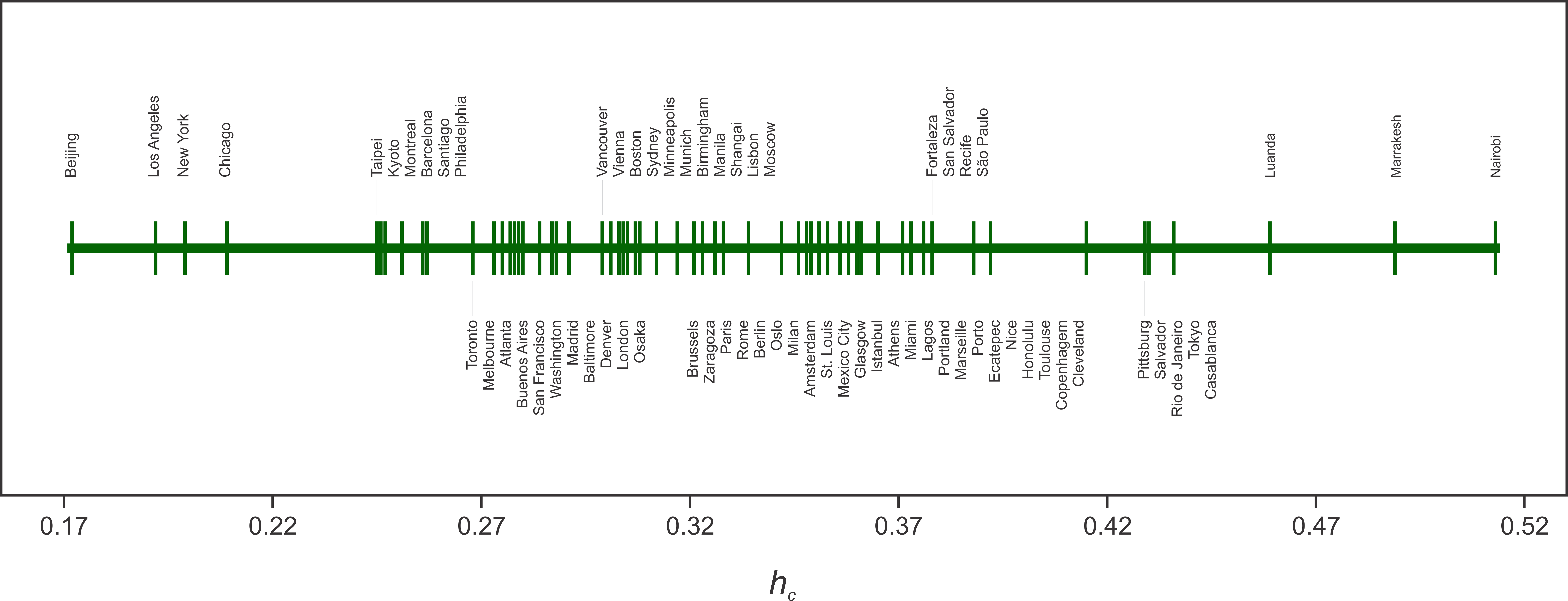}
\includegraphics[width=0.45\textwidth, angle=0]{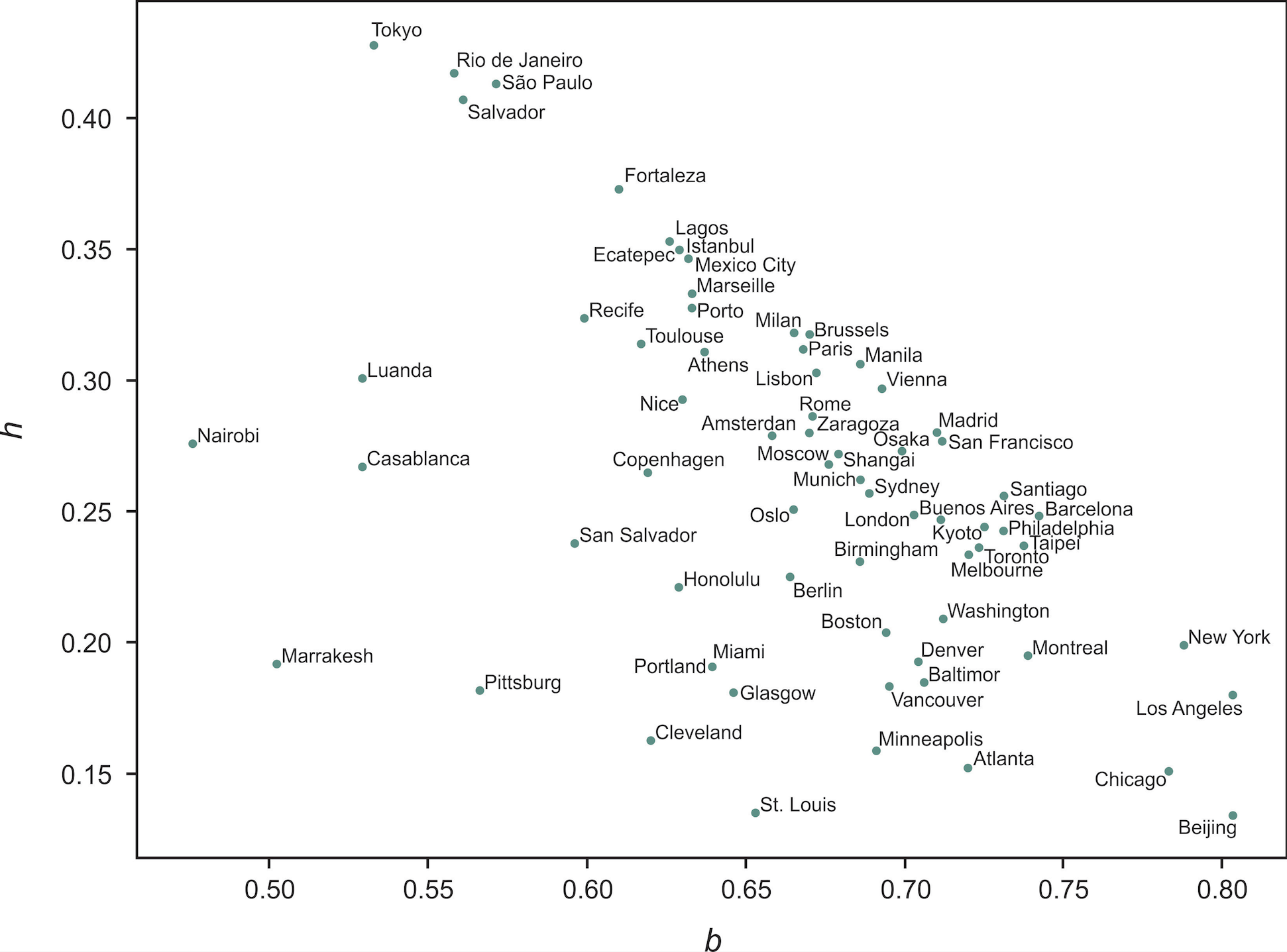}
\includegraphics[width=0.45\textwidth, angle=0]{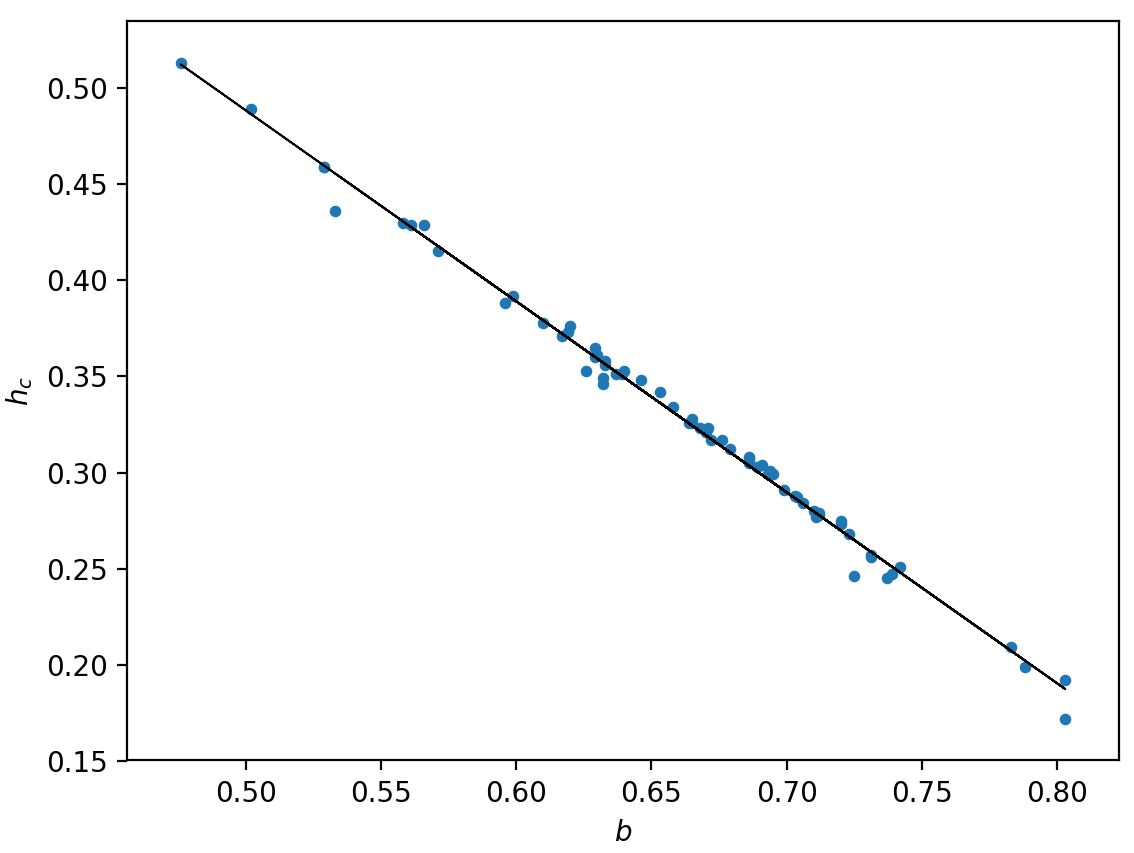}
\caption{ 
{\it Top:} Estimated values of $h_c$ 
for the cities under analysis.
{\it Bottom: } On the left, scattering plot of the $h$ values versus $b$ for all the considered cities. 
On the right, $h_c$ versus $b$. In this case, data present a Pearson's correlation equal to $-0.997$
and are fitted by the suggestive linear relation: $y=-0.99 x + 0.99$. This fact implies that
the two fitting parameters  $c$ and $b$ of equation \ref{fitting} are not independent, but $c\approx b$.
}
\label{fig_entroCorr}
\end{figure}

In Figure \ref{fig_entroCorr} we display on a horizontal axis the results for the estimation of the corrected entropy $h_c$ for the sampled cities. This measure introduces a clear sorting among our data.
It is interesting to compare the behaviour of the corrected and the original entropy $h$ with
the values of the parameter $b$ of the function \ref{fitting}, as obtained from fitting
the mean values of $H_n/n$. 
In fact, this parameter characterise how the block entropy estimates $H_n/n$ converge to the asymptotic values $h$. There are no clear theoretical interpretation of this quantity, but it can be generally related to the presence of long range correlations. As can be seen form the scattering plot of $h$ versus $b$, there is not an 
evident relation between these two measures. 
In contrast, once corrected by the factor $1-H_1$, the two quantities anticorrelate almost perfectly:
low values of $h_c$ correspond to high $b$ values, following a linear dependence.
This is consonant with the fact that $b$ is determined by the rate with which correlations at 
different scales decay, and $h_c$ measures the randomness which survives to correlations, 
not containing the asymmetry in the distribution of the symbols 0 and 1 of our maps.
Furthermore, it is interesting to note that the values of the parameters $b$ are contained in the interval [0.48, 0.80]. These values are consistent with the entropy convergence found in written texts, where $b$ ranges from 0.4 to 0.6 \cite{Ebeling,Ebeling2}, and with a result for a Beethoven sonata where an exponent 0.75 was found \cite{Anishchenko}. 
These results seem typical of language-like systems, where the presence of long-range order is characterised by a slowly decaying contribution to the asymptotics of the entropy for large $n$. 

\begin{figure}[h]
\centering
\includegraphics[width=0.7\textwidth, angle=0]{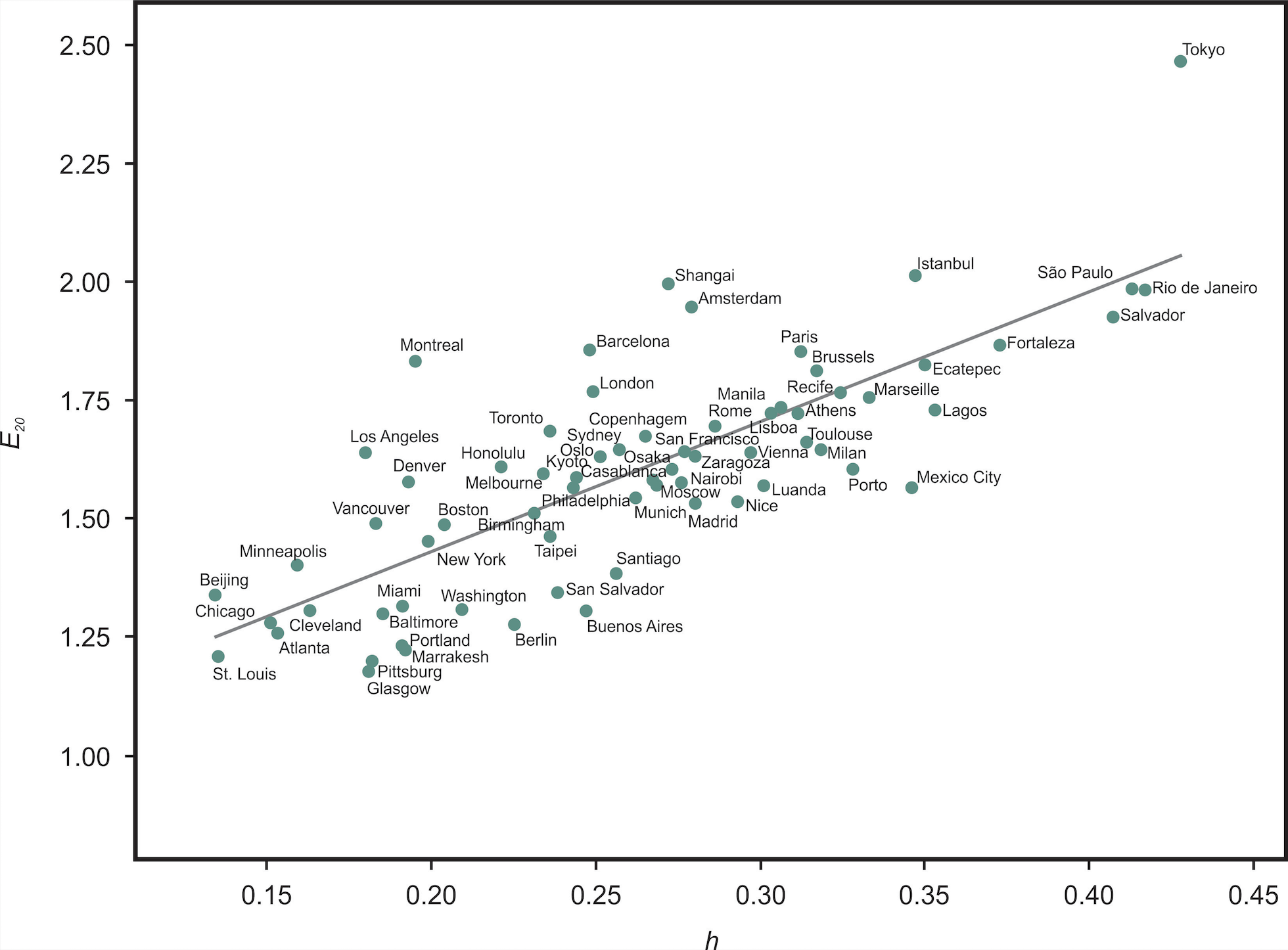}
\caption{ 
Estimated values of $h$ and excess entropy $E_{20}$ for cities under analysis. 
}
\label{fig_trajectory3}
\end{figure}

The estimated values of $b$ correspond to a slow convergence, which  implies  that the excess entropy
is infinite. In fact, the excess entropy can be alternative defined as the subextensive part of $H_n$:
$E=lim_{n \to \infty}[H_n-h\cdot n]$ \cite{Crutchfield_Chaos03}. 
As for our systems $H_n\sim n^{\gamma}+h\cdot n$,  the excess entropy $E$ diverges as a power-law, 
with $\gamma=1-b$.
For this reason, we estimate the n-finite excess entropy $E_M$ for $M=20$.
In Figure \ref{fig_entroCorr} we plot ($h$, $E_{20}$) pairs. 
This plot is a Complexity-Entropy diagram \cite{Crutchfield_PRL89,Feldman_Chaos08}
which displays how the n-finite excess entropy and the entropy are related.  
Our data points are sparse and relatively scattered, nevertheless
they suggest a monotone and linear relation between the two quantities.
The maximum of the finite excess entropy corresponds to 
the maximum of randomness, as expressed by $h$. 
This is possible because, in general,  data which presents high $h$ values, also
present slow convergence of the finite entropy  estimates $h_n$ to their asymptotic value.
These results suggest that the excess entropy is not able to introduce a classification in our dataset different 
from the one generated by the entropy $h$: structures and paths are not clearly detected 
as different from naive disorder. 
In particular, it presents the same drawbacks of $h$, classifying with a low level of
statistical complexity cities which present a high asymmetry in the ratio
built/unbilt regions but that, from a visual inspection, clearly
present a high level of structures, paths or randomness, like, for example, Marrakesh, Casablanca and Nairobi.
Moreover, as our estimation, from a theoretical point of view, 
does not correspond to the excess entropy, but to the n-finite excess entropy, 
we consider that this quantity fails in generating 
the wished alternative classification based on a complexity measure.

\begin{figure}[h]
\centering
\includegraphics[width=0.9\textwidth, angle=0]{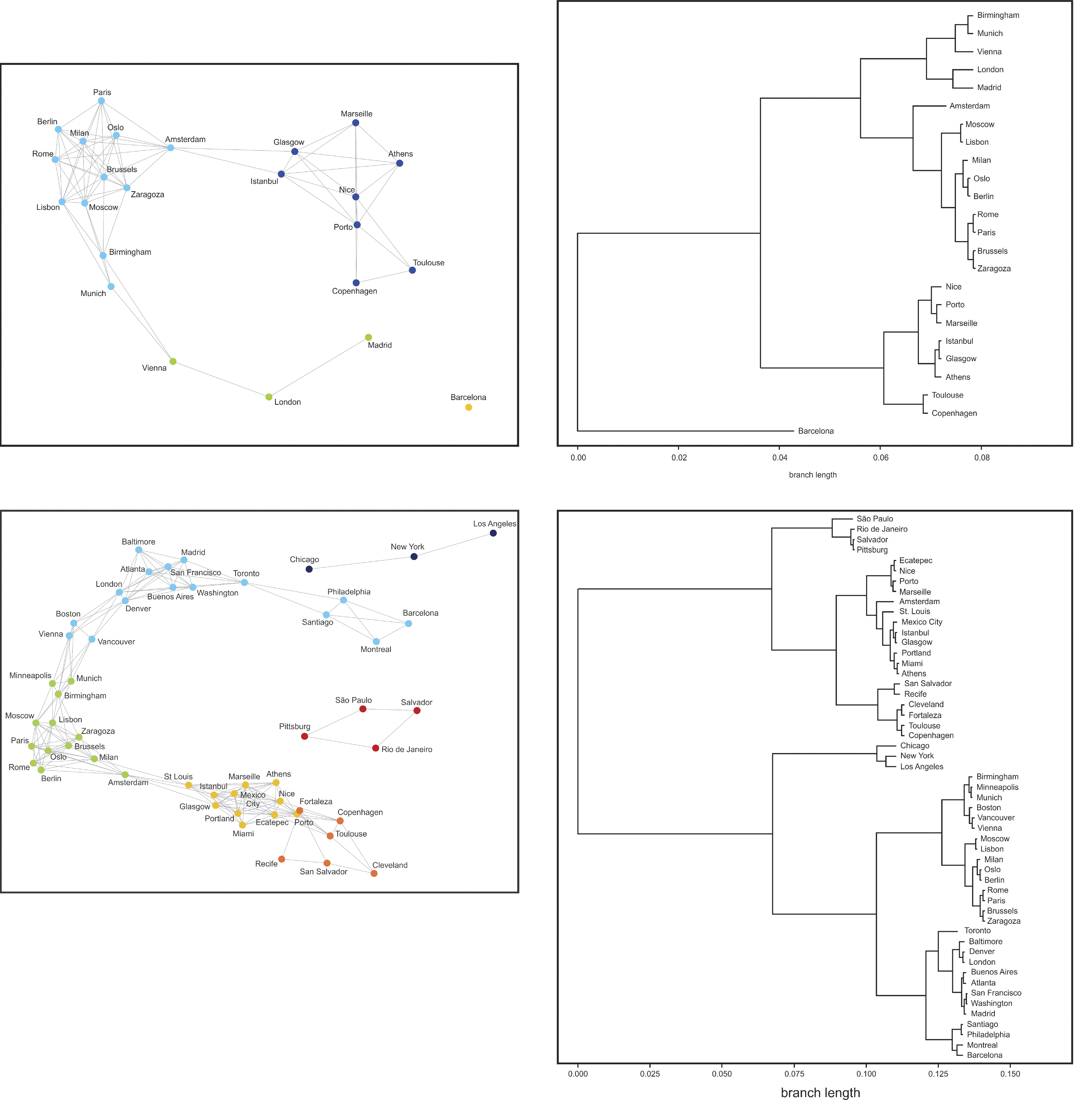}
\caption{ {\it Top:} On the left, the proximity network of of the considered European cities based on the value of $h_c$. The edge lengths are not proportional to the considered distances between cities. The node colour correspond to different communities. 
On the right, dendrogram. 
{\it Bottom:} Proximity network and dendrogram of the analysed European, North and Latin American cities.
}
\label{fig_EuAmer}
\end{figure}

For this reason we develop our classification scheme by using the $h_c$ values.
The similarity networks are constructed fixing the threshold value to $0.015$, which corresponds to the 80\% confidence interval of the extrapolated values of $h_c$. 
We implement the clustering analysis in increasing subsets of our pool of cities, starting within a same region. This way, it is easier to extract and visualise potential patterns or clusters of cities sharing similar entropy levels.
We started by looking into European cities (figure \ref{fig_EuAmer}). As can be seen from
the community detection results, selected cities in Europe cluster in 4 groups in the proximity network. 
Similar results are displayed by the corresponding dendrogram.

Next, we analysed the cities of Europe and the Americas (figure \ref{fig_EuAmer}). 
We can distinguish 6 different clusters in the proximity network. 
Results show low entropy clusters, particularly the one with Chicago, New York and Los Angeles, followed by clusters of increasing entropy values. Interestingly, Montreal, Philadelphia, Santiago and Barcelona are clustered, as areas under analysis in these cities were historically structured in rigid grid-like patterns. Others European cities appear in the following clusters, mostly mixing up cities from different regions. More discernible regions emerge again at the bottom, as most Latin American cities are concentrated in higher entropy levels: Recife (Brazil) and San Salvador (El Salvador, Central America), and Sao Paulo, Rio de Janeiro and Salvador in Brazil.

\begin{figure}[h]
\centering
\includegraphics[width=0.9\textwidth, angle=0]{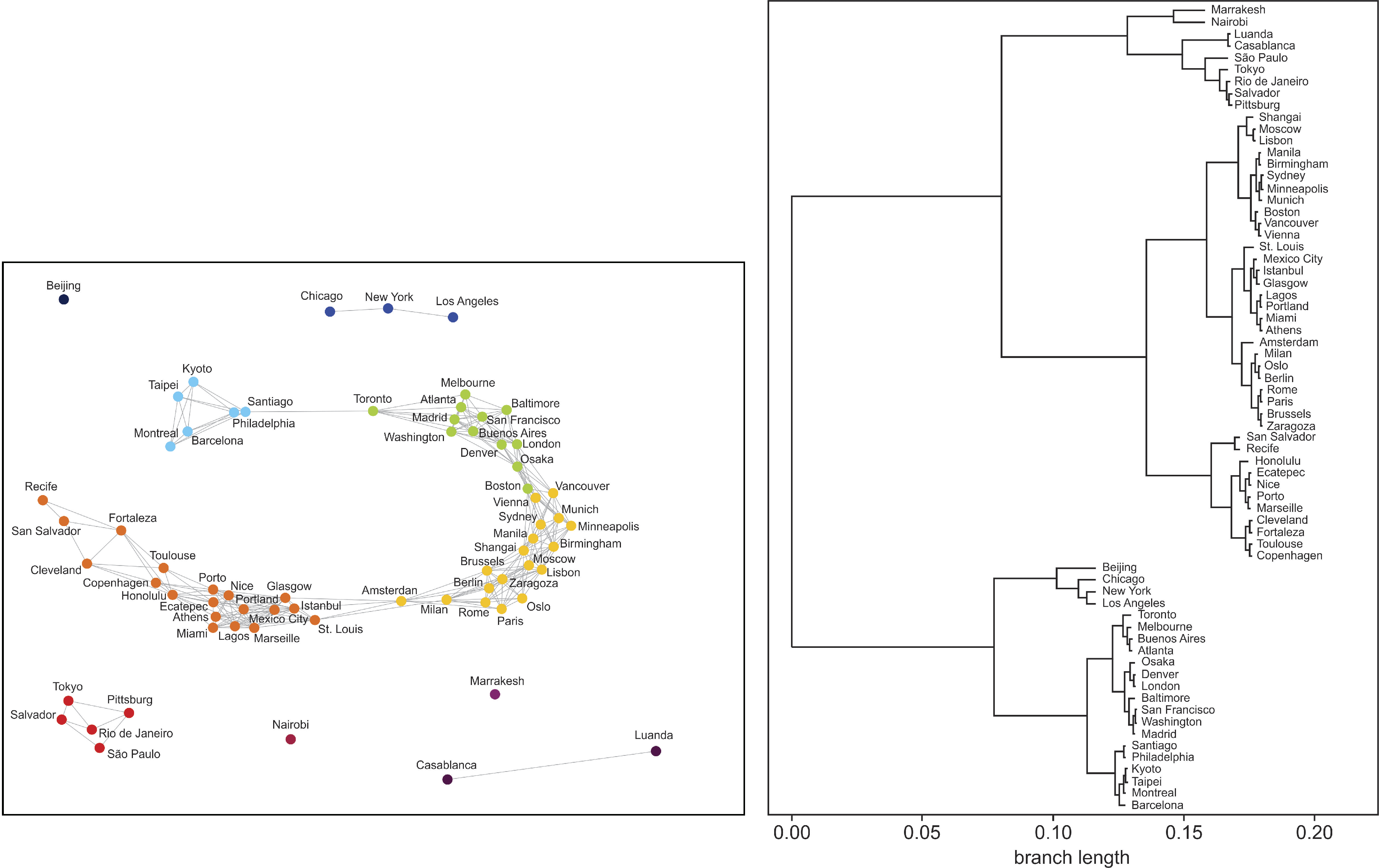}
\caption{
Proximity network and dendrogram of European, American, African, Asian and Oceanian cities under analysis.}
\label{fig_EuAmerAfrAsi}
\end{figure}

The concluding analysis joins together all the considered cities, adding the Asian and Oceanian data. The number of clusters in the proximity network is comparable to the former analysis. The community structure is
similar to the previous one, with the Asian cities distributed among the pre-existing clusters. 
The complete dendrogram can be seen in figure \ref{fig_EuAmerAfrAsi}.
Looking at the branch length around 0.075, the dendrogram shows two major branches. 
The one on top includes cities of high entropy values, and it further bifurcates in specific groups: Marrakesh (Morocco) and Nairobi (Kenya); Luanda (Angola) and Casablanca (Morocco), cities in Northern and Central Africa; S\~ao Paulo (Brazil) and Tokyo (Japan), two massive cities with similarities in their fragmented cityscapes; Rio de Janeiro and Salvador (Brazil) and Pittsburg (US) - the latter, a remarkable exception in the US scenario characterised by more ordered built form systems. 
These clusters are also visible in the proximity network analysis. This high end of the entropy spectrum ($h_c>0.38$) is where African and most Latin American cities can be found. The four cities with the highest entropy levels are in Africa. Casablanca and Marrakesh in Morocco, an Islamic country, along with Nairobi and Luanda, display the highest entropy levels ($h_c>0.43$). Brazilian cities in South America and San Salvador in Central America. Fortaleza, Recife, S\~ao Paulo, Salvador and Rio de Janeiro, five Brazilian cities analysed, are consistently found with similarly high entropy levels ($0.38<h_c<0.43$). Tokyo and Pittsburg are exceptions to these regional trends, with entropy values quite different from other cities examined in their respective regions.

This main branch ramifies into larger groups, arranged along two other branches in the dendrogram (around branch length 0.15). They are also visible in the proximity network and comprise cities from different regions, with 
$0.30 < h_c < 0.40$.
At larger branch length other bifurcations emerge. We find Shanghai, Moscow and Lisbon; Manila and Birmingham; Sydney, Minneapolis and Munich; Boston, Vancouver and Vienna. At a similar branch length, we find St. Louis; Mexico City; Istanbul and Glasgow; Lagos, Portland, Miami and Athens. Then we have Amsterdam; Milan, Oslo and Berlin; Rome, Paris, Brussels and Zaragoza in ramifications in the dendrogram and, accordingly, in close positions in the proximity network. A different ramification 
includes San Salvador and Recife; Honolulu; Ecatepec and Nice, Porto and Marseille; and Cleveland and Fortaleza, Toulouse and Copenhagen. In turn, the proximity networks analysis groups these cities into two major clusters, with their constellations of positions following the relations detailed in the dendrogram. This is a rather diverse set of cities. Nevertheless, we can see some internal regional consistencies. For instance, most European cities are found in these clusters. We can also identify that Southern European cities frequently find slightly higher entropy levels (around 0.37). But there are no clear differences here since cities from different regions share some of the same branches and clusters and find similar entropy levels.

The second major branch , emerging at length around 0.075, 
comprises cities closer to the other end of the spectrum. Low entropy clusters in our sample are more easily discernible and relatable to geographical regions and potentially linked 
specific spatial cultures characterised by higher order levels. We find here Beijing, a city planned under rigorous rules since the XIV century, in a detached position both in the dendrogram and in the proximity network. We have Chicago, New York, and Los Angeles 
which are also closely linked in the network. At length 0.12 in the dendrogram, we have more bifurcations and smaller branches: Toronto; Melbourne, Buenos Aires and Atlanta; Osaka, Denver and London; Baltimore, San Francisco, Washington and Madrid. Another smaller low entropy branch 
includes Santiago and Philadelphia; Kyoto, Taipei and Montreal; and Barcelona. These cities also share a cluster in the proximity network analysis. In turn, similarities in entropy values (figure 2) shed more light on regional consistencies. 

\section{Discussion}

Our analysis shows that the Shannon entropy estimation of the maps is reliable and robust.
An exiting by-product of this analysis reveals that 
our typical entropy convergence behaviours are 
similar to the ones found in written texts and sonatas, characterised by a slow convergence toward
the asymptotic entropy values.
An interesting working hypothesis suggested that strings characterised by this slow decay, being on the borderline between order and chaos, might be prototypes of information carrying sequences \cite{Ebeling91}.
This result, which shows  a parallel between the human shaped urban form and language-like systems, 
reinforces  the insight of looking at the 
 built form as an information carrying substrate and 
at human activities as diverse as building, writing and composing, as a complex process generating forms characterised by subtle, involved correlations at different scales which support entangled long-range structures.

From the Shannon entropy of the maps we introduce the corrected entropy $h_c$ which can perform very well
for constructing a classification scheme which distinguishes similarity anchored on the measurement of
randomness in built morphology. 
The results of the clustering analysis
display a mix of similarities between cities within the same region and some intricacies,
as cities from different regions also may cluster around similar entropy values. Furthermore, a same country may have cities of completely different entropy levels. For instance, we find Kyoto in Japan a highly ordered city, whereas entropy levels in Tokyo are among the highest in our sample. In the US, the overall trend is towards the lowest built form entropy levels examined, but Pittsburg's entropy is more similar to those generally found in Brazilian cities. 

What are the reasons for such regional patterns and variations in clusters? Firstly, many cities from the same regions converge around similar entropy values. This finding suggests that cities within regions share specific ways of ordering space and built form, consistent with the hypothesis of regionally bounded spatial cultures. However, if entropy values may characterise the region, they do not do so exclusively. Many cities from different regions in our sample showed similar levels of order in their built form systems. 
So what could explain such similarities across different regions? Are they mere coincidence? They might be. Different spatial cultures may order space into different shapes, but these shapes might contain similar order levels. As their populations interact across distances, they also might influence one another in time. Furthermore, they might be subject to geographical contingencies. In short, they might be subject to specific morphogenetic forces shaping their spatial arrangements. 

This problem is defined in urban studies as ``morphogenesis", the production of urban spaces and the creation of built form. It leads to the emergence of patterns shaped by forces like social organisation forms and ordering space based on different emphases on order. A proper look into morphogenesis means looking into the histories of these cities, something beyond the scope of this paper. However, there is literature - not a large one - focused on morphogenetic processes. Since the mid-1960s, it has concentrated on sets of forces known 
in complexity science 
as bottom-up and top-down \cite{alexander1964notes,batty2013}. Authors from both historical \cite{kostof1991city} and configurational perspectives \cite{hillier1984social,battyetal1989} see bottom-up forces as self-organisation processes. These processes involve the daily interactions of people, progressively producing buildings and open spaces that amount to the dense systems we call cities. They include people's conscious and unconscious ways of spatially organising their interactions, along with cultural emphases on order. In turn, top-down processes are usually seen as conscious efforts performed by specialised agencies created precisely to control (supposedly) messy bottom-up processes. These agencies do so for different reasons: to avoid conflicts between actions or interests, like social groups or classes, to avoid unintended large scale consequences of micro-scale individual actions \cite{schelling2006micromotives}, and so on. They may do this through different means, like through institutions to evaluate urban performance or rules to guide urban growth. 

These practices materialised over time as a field, urban planning. Planning may interfere with morphogenetic processes when it tries to dictate how the built form should be produced and spatially arranged. Planning views, like any idea, may spread in space, transcending regions. They also shift in time, focusing on different aspects - say, from strict definitions on how building ensembles should be arranged to local rules focused on parts of the urban system, like streets, plots, buildings, etc. Furthermore, such bottom-up and top-down forces are frequently active at the same time. An accurate analysis of this interplay requires a mix of historical description of urban evolution and the spatial analysis of pattern formation applied to cities and regions. We cannot offer this holistic approach here, but we can provide an initial framework to look into this interplay. There is a field of different emphases on planning with potential morphological implications, which include: 
	(a) Order-oriented, top-down control in planning rules, like in North American cities, Beijing, Barcelona or Santiago, leading to low entropy levels in built form. 
	(b) Historically emergent patterns geared by urban traditions focused on continuity and alignment in buildings and facades strung along streets and topographic lines, like Vienna, London and Madrid, despite more organic street networks. Such cities tend to have slightly higher entropy levels than in (a). 
	(c) Top-down planning cultures based on piecemeal definitions of urban areas and plots under rules, along with regulations that do not specify the position of buildings in their plots (e.g. frontal and lateral setbacks) and neighbouring buildings. Such cities may exhibit high built form entropy.  
	(d) Bottom-up emergent patterns based on highly variable built form and street networks leading to high levels of disorder and entropy, like the case in urban areas and informal settlements, particularly in developing countries like Morocco. These features are also likely to lead to high entropy levels in built form.

These possibilities are a non-exclusive set of morphological paths related to different levels and forms of planning, of course. This set is helpful to discern spatial cultures that transcend regions and find similarities and differences in planning forms. In our case, clusters in the high and low ends of the entropy spectrum are more easily discernible and potentially relatable to specific planning cultures. Indeed, most high entropy cities in our sample are found in developing regions, like Latin America and Africa. Cities with more irregular physical patterns are usually thought to result from development left entirely to individuals, as bottom-up processes lead to the unplanned city's random ways. But this is not necessarily the case. Cities in countries like Brazil or Nigeria seem particularly subject to trends (c) and (d), sharing emphases on local rules focused exclusively on individual buildings rather than coordinated construction. This finding brings no aesthetic or moral judgement. Parcel-based, piecemeal developments exempt from requirements to keep connections to neighbouring areas, including street continuity and grid alignment among nearest neighbours, can easily lead to a high fragmentation level. Latin American cities Buenos Aires' and Santiago's central areas are exceptions to this trend, as they were founded in the Sixteenth century by Spanish colonisers following rigid orthogonal patterns. 

In turn, top-down processes triggered by governing agencies guide the organisation of urban land and built form, leading to more uniformly ordered cities \cite{kostof1991city}  and low entropy configurations. Chicago, New York and Los Angeles epitomise that trend. The North American tradition in urbanism is based on orthogonal grids, with great regularity in urban blocks. So does Beijing in China, with its planning based on `regulations of construction' along with cardinal directions following a tradition since the early Ming dynasty (1368-1644 AD). In Europe, control over built form is emblematically found in Barcelona. Most of Barcelona's section analysed in our sample was strictly defined in 1859 by Ildefons Cerd\'a's Eixample plan fixing block systems and systematically continuous facades. In turn, major cities like Vienna, London and Madrid display highly consistent built forms regarding fa\c cade alignment and continuity in urban blocks, despite their organic street networks with 
changing street orientations.

The 
clustering analysis shown above suggests that we cannot associate particular entropy levels exclusively with specific regions of the world. This finding suggests a few possibilities, not mutually exclusive: (i) These cities may not have reached distinctive features enough to be captured by a measure geared to assess randomness in cellular arrangements of built form. (ii) These cities have not reached enough differences in their cellular arrangements analysed in the 
areas selected. (iii) The measure may not be precise enough to capture every 
morphological difference between cities. 
Cultural idiosyncrasies might be encoded in the built form, but they do not display sufficient differences in randomness at the spatial level analysed. (iv)  Different spatial arrangements might reach similar entropy levels. 

That said, our findings also show that cities within specific regions do tend to converge around similar entropy values, suggesting that they share common built form features and, most importantly, share ways of ordering space in their morphogenesis. Our measure is useful for capturing spatial information related to different emphases on order and coordination latent in these regions and grasping similarities between planning cultures across different regions.

\ack

V.M.N. wants to thank CNPq for financial support (research project number 315086/2020-3).
F.N.M.S.F. received partial financial support from the PIBIC program of Universidade Federal do Rio de Janeiro.

\end{document}